\documentclass[floatfix,showpacs,amsmath,amsfonts,amssymb,aps,twocolumn,longbibliography,superscriptaddress,prb]{revtex4-2}

\usepackage[table,dvipsnames]{xcolor}
\usepackage{graphicx}
\usepackage{dcolumn}
\usepackage{amsthm}
\usepackage{bm}
\usepackage{siunitx}
\usepackage{nicefrac}
\usepackage{mathtools}
\usepackage{float}
\usepackage{braket}
\usepackage{makecell} 
\usepackage{multirow}
\usepackage{wasysym}
\usepackage{titletoc}
\usepackage{multirow}
\usepackage{mwe}
\usepackage{diagbox}
\usepackage{pifont}

\definecolor{darkred}{rgb}{0.6,0.,0.}
\definecolor{darkgreen}{rgb}{0.,0.5,0.}
\definecolor{darkblue}{rgb}{0.,0.,0.6}
\usepackage[breaklinks, colorlinks=true, linkcolor=darkred, citecolor=darkgreen, urlcolor=darkblue]{hyperref}

\expandafter\def\csname ver@etex.sty\endcsname{3000/12/31}

\newcommand{\includegraphicsgood}{\includegraphics}

\usepackage[inactive,tightpage]{preview}
\PreviewMacro[*[[!]{\input{}}
\PreviewMacro[*[[!]{\includegraphicsgood}
\newcommand{\normord}[1]{:\mathrel{#1}:}

\begin{document}

\title{Stability of fractional Chern insulators with a non-Landau level continuum limit}
\author{Bartholomew Andrews}
\affiliation{Department of Physics, University of California at Berkeley, 100 South Dr, Berkeley, California 94720, USA}
\affiliation{Department of Physics and Astronomy, University of California at Los Angeles, 475 Portola Plaza, Los Angeles, California 90095, USA}

\author{Mathi Raja}
\affiliation{Department of Physics, University of California at Berkeley, 100 South Dr, Berkeley, California 94720, USA}

\author{Nimit Mishra}
\affiliation{Department of Physics and Astronomy, University of California at Los Angeles, 475 Portola Plaza, Los Angeles, California 90095, USA}

\author{Michael P. Zaletel}
\affiliation{Department of Physics, University of California at Berkeley, 100 South Dr, Berkeley, California 94720, USA}

\author{Rahul Roy}
\affiliation{Department of Physics and Astronomy, University of California at Los Angeles, 475 Portola Plaza, Los Angeles, California 90095, USA}

\date{\today}

\begin{abstract}

The stability of fractional Chern insulators is widely believed to be predicted by the resemblance of their single-particle spectra to Landau levels. We investigate the scope of this geometric stability hypothesis by analyzing the stability of a set of fractional Chern insulators that explicitly do not have a Landau level continuum limit. By computing the many-body spectra of Laughlin states in a generalized Hofstadter model, we analyze the relationship between single-particle metrics, such as trace inequality saturation, and many-body metrics, such as the magnitude of the many-body and entanglement gaps. We show numerically that the geometric stability hypothesis holds for Chern bands that are not continuously connected to Landau levels, as well as conventional Chern bands, albeit often requiring larger system sizes to converge for these configurations.

\end{abstract}

\maketitle         

\section{Introduction}
\label{sec:intro}

For over 40 years, the fractional quantum Hall effect (FQHE) has inspired a large and diverse body of research, since it is the original example of a topological phase of matter with fractional excitations and has the scope for revolutionary applications in quantum metrology and computing~\cite{vonKlitzing20, Nayak08}. In the past decade, research interest has focused particularly on lattice generalizations of these fractional quantum Hall states, known as fractional Chern insulators (FCIs)~\cite{Liu23, Regnault11, Moller09}, which enrich FQHE physics through their enhanced configurability and have the potential to be realized at zero magnetic field~\cite{Neupert11, Sheng11} and high temperatures~\cite{Tang11}. Most notably, the past few years have seen exciting experimental reports of low- and zero-field FCIs in moiré lattices of graphene~\cite{Xie21, Lu24} and transition-metal dichalcogenides~\cite{Cai23}, which has reignited theoretical interest in FCI stability~\cite{Ledwith20, Wang22, Mera21, Bouhon23, Fujimoto24, Duran23, Dong23_2, Guo23, Abouelkomsan23, Wang23}. The underlying goal of such research is to use the single-particle band structure to predict properties of the many-body spectrum, in order to identify so-called ``ideal" Chern bands that yield the most robust FCIs~\cite{Wang21, Ledwith22, Guan22}. The cornerstone of FCI stability theory is the geometric stability hypothesis~\cite{Jackson15}, which conjectures that FCIs become more robust as the trace inequality is saturated, on the basis that at this point the Chern band projected density algebra is identical to that of Landau levels~\cite{Roy14, Parameswaran13}. This hypothesis has been rigorously tested in a variety of topological flat band models and is found to hold in many typical cases~\cite{Jackson15, Bauer16, Bauer22}. However, recently, several extensions to the geometric stability hypothesis have been proposed~\cite{Simon20, Ledwith22, Estienne23, Fujimoto24, Abouelkomsan23}, with authors arguing analytically that the single-particle resemblance to Landau levels cannot be the defining criterion of FCI stability but rather, a special case. In light of this, there is motivation to construct a simple many-body lattice model that can numerically test the geometric stability hypothesis decoupled from the Landau level limit.

In this paper, we investigate the scope of the geometric stability hypothesis by studying FCIs in a family of topological tight-binding models that do not have a Landau level continuum limit. Specifically, by adding a set of longer-range hoppings to the Hofstadter model, we show that terms quadratic in momentum exactly cancel in its small flux density expansion, which breaks the $SO(2)$ symmetry present in the Landau level Hamiltonian. In typical topological flat band models, the continuum and Landau level limits are synonymous and so the effect of each is obscured, whereas in our model we can decouple these limits, and approach non-Landau flat bands in the continuum. This is particularly useful for probing the geometric stability hypothesis, which explains the generally increasing stability of FCIs in the continuum limit, as a consequence of the increasing similarity of their host Chern bands to Landau levels. In this framework, we study the many-body properties of bosonic and fermionic Laughlin states numerically using exact diagonalization on a torus, and we quantify the robustness of FCIs using many-body gaps in their energy and entanglement spectra. We show how the geometric stability hypothesis may appear to break down as we approach the non-Landau level continuum limit, at system sizes and flux densities typically used in exact numerics~\cite{Jackson15, Bauer16, Bauer22}, however is recovered as we take the thermodynamic or continuum limits, which highlights the limitations of comparing FCI stability using finite-size numerics. We argue that the geometric stability hypothesis is widely supported by numerics for small systems because the continuum and Landau level limits typically coincide~\cite{Jackson15, Bauer16}. Moreover, we comment on proposals that the underlying FCI stability criterion goes beyond the resemblance of single-particle spectra to Landau levels~\cite{Simon20, Ledwith22, Estienne23}.

The structure of this paper is as follows. In Sec.~\ref{sec:model}, we introduce the single-particle and many-body Hamiltonians, and we explain our choice of system parameters. In Sec.~\ref{sec:method}, we outline the method for quantifying the geometric stability hypothesis, as well as the stability of FCIs. In Sec.~\ref{sec:results}, we present the exact diagonalization results for the bosonic and fermionic Laughlin states, and compare properties of the single-particle and many-body spectra. Finally, in Sec.~\ref{sec:conc}, we summarize and discuss the results, and comment on their implications.

\section{Model}
\label{sec:model}

In this section, we introduce the lattice model. In Sec.~\ref{subsec:sp}, we focus on the single-particle Hamiltonian, and in Sec.~\ref{subsec:mb}, we include interactions to define the complete many-body model.

\subsection{Single-particle model}
\label{subsec:sp}

We consider a system of $N$ particles hopping in the $xy$-plane, on a finite square lattice with basis vectors $\{\mathbf{a}_x, \mathbf{a}_y\}=a\{\hat{\mathbf{e}}_x, \hat{\mathbf{e}}_y\}$, spacing $a$, and dimensions $N_x \times N_y$, with periodic boundary conditions, in the presence of a perpendicular magnetic field $\mathbf{B}=B\hat{\mathbf{e}}_z$. We define the single-particle Hamiltonian of this system as the generalized Hofstadter model~\cite{Harper55, Azbel64, Hofstadter76}
\begin{equation}
\label{eq:sp}
H_0 = -\sum_{n>0} \sum_{\braket{ij}_n} t_n e^{\mathrm{i}\theta_{ij}} c^\dagger_i c_j + \text{H.c.},
\end{equation}
where $\braket{\dots}_n$ denotes $n$th nearest-neighbors (NN) with corresponding hopping strength $t_n$, $\theta_{ij}$ denotes the Peierls phase acquired by hopping from site $i$ to site $j$, and $c^{(\dagger)}_i$ is the particle (creation)annihilation operator at site $i$. In particular, the Peierls phases are defined as $\theta_{ij}=(2\pi/\phi_0)\int_i^j \mathbf{A}\cdot\mathrm{d}\mathbf{l}$, where $\phi_0=h/e$ is the flux quantum, $\mathbf{A}$ is the vector potential, and $\mathrm{d}\mathbf{l}$ is the infinitesimal line element going from site $i$ to site $j$. These phases incorporate the perpendicular magnetic field in the system to a good approximation~\cite{Peierls33, Zak64}, and they define a magnetic unit cell (MUC) of dimensions $l_x \times l_y = q$, such that $L_x\times L_y$ are the system dimensions in MUC units. The flux density of the system is then given as $n_\phi=Ba^2/\phi_0=p/q$, where $(p,q)$ are coprime integers, and $q\to\infty$ corresponds to the continuum limit. The incommensurability between the MUC area and the flux quantum gives rise to the celebrated fractal energy spectrum as a function of $n_\phi$, known as the Hofstadter butterfly~\cite{Hofstadter76, Osadchy01, HofstadterTools}. Consequently, the Hofstadter model is a popular choice, for both theorists~\cite{Andrews18, Andrews21, Andrews21_2, Moller15, Motruk16} and experimentalists~\cite{Aidelsburger13, Dean13, Aidelsburger15, Lu21, Yu22}, due to its ability to generate an infinite selection of topological band structures with arbitrary Chern number.

In the conventional Hofstadter model (with $t_{n>1}=0$), it is straightforward to show that the lattice model approaches the Landau level Hamiltonian in the continuum limit. We first write the NN Hofstadter model, $H_\mathrm{NN}$, as a symmetric sum of magnetic translation operators, such that
\begin{equation}
H_\mathrm{NN} = -t_1 (T_x + T_y) + \mathrm{H.c.},
\end{equation}
where $T_m=\sum_{\braket{ij}_{1,m}} e^{\mathrm{i}\theta_{ij}}c^\dagger_i c_j$ and $\braket{ij}_{1,m}$ denotes 1st-NN in the $\hat{\mathbf{e}}_m$ direction~\footnote{The symmetric sum is a gauge choice.}. Since magnetic translation operators are unitary, we can express them in terms of the Hermitian generators $T_m = e^{\mathrm{i}K_m}$, where the pseudomomenta $K_m$ are the lattice analogues of the dynamical momenta $\pi_m$ and satisfy $[K_x, K_y]=2\pi n_\phi \mathrm{i}$. This yields
\begin{equation}
H_\mathrm{NN} = -2 t_1 (\cos K_x + \cos K_y).
\end{equation}
Finally, we can make the dependence on the flux density explicit by defining $P_m = n_\phi^{-1/2} K_m$ and expand in small $n_\phi$ to yield
\begin{equation}
H_\mathrm{NN} = -4 + n_\phi (P_x^2 + P_y^2) + O(n_\phi^2).
\end{equation}
Using the fact that $P_m=(\ell/\hbar)\pi_m$, where $\ell$ is the magnetic length, we can show that the effective Hamiltonian $H_\mathrm{NN} \sim n_\phi (P_x^2 + P_y^2)$ is isomorphic to the Landau level Hamiltonian $H_\mathrm{LL} = (\pi_x^2 + \pi_y^2)/(2m^*)$ with $m^*=\hbar^2/(2a^2)$. Further details of this derivation are discussed in Ref.~\onlinecite{Bauer22}.

In the generalized Hofstadter model (with $t_{n>1}\neq0$), the procedure for taking the continuum limit follows in a similar way. For simplicity, we restrict ourselves to the generalized Hofstadter model with hoppings along the cardinal axes, $H_{\perp}$, to avoid cross terms in our expansion. We start by writing the model as a symmetric sum of magnetic translation operators
\begin{equation}
H_\perp = -\sum_{n>0} t'_n (T_x^n + T_y^n) + \text{H.c.}, 
\end{equation}
where $\{t'_1, t'_2, t'_3, t'_4, \dots\}=\{t_1, t_3, t_6, t_9, \dots\}$ is the set of hoppings along the cardinal axes. Next, we write the magnetic translation operators in terms of the Hermitian generators $T_m=e^{\mathrm{i}K_m}$, introduce $P_m = n_\phi^{-1/2} K_m$, and expand in small flux density, which yields
\begin{equation}
\begin{split}
H_\perp = -4\sum_{n>0} t'_n &+ n_\phi (P_x^2 + P_y^2) \sum_{n>0} n^2 t'_n \\
&- n^2_\phi (P_x^4 + P_y^4) \sum_{n>0} \frac{n^4}{12} t'_n \\
&+ n^3_\phi (P_x^6 + P_y^6) \sum_{n>0} \frac{n^6}{360} t'_n + O(n_\phi^4).
\end{split}
\end{equation}
Crucially, by including longer-range hoppings in the Hamiltonian, we are able to tune the coefficients in our expansion and depart from the Landau level limit. 

\begin{figure}
	\includegraphics[width=.8\linewidth]{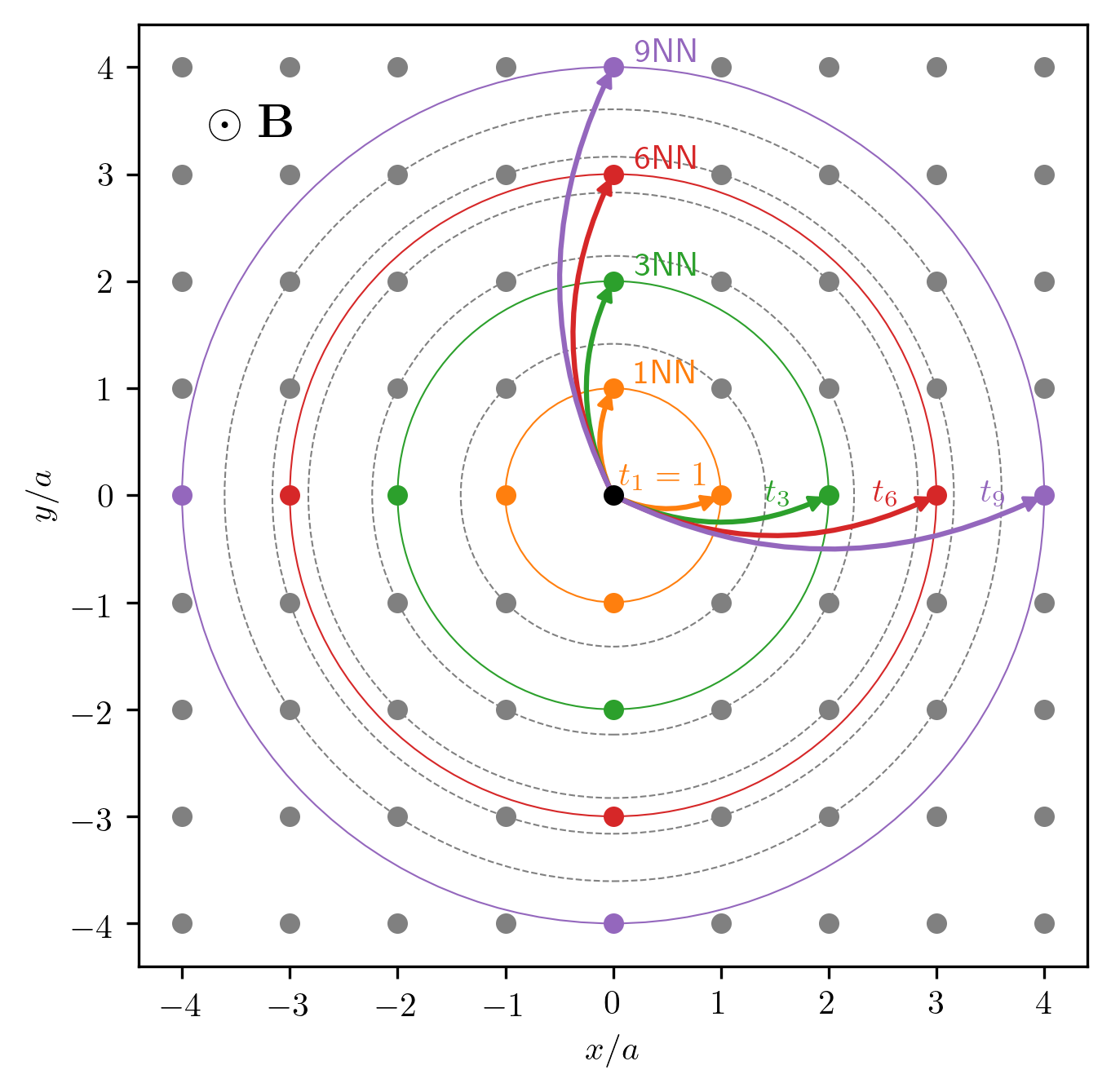}
	\caption{\label{fig:model} \textbf{$(3,6,9)$ Hofstadter model.} Sketch of the hopping terms in the $(3,6,9)$ Hofstadter Hamiltonian (conjugates not drawn). The radii for the first 9 sets of nearest neighbors on a square lattice are: $1$, $\sqrt{2}$, $2$, $\sqrt{5}$, $2\sqrt{2}$, $3$, $\sqrt{10}$, $\sqrt{13}$, $4$.}
\end{figure}

In this paper, we focus on the generalized Hofstadter model with $n\in\{1,3,6,9\}$ hoppings along the cardinal axes and $t_1=1$, as depicted in Fig.~\ref{fig:model}, which we refer to simply as the $(3, 6, 9)$ Hofstadter model, $H_{(3,6,9)}$. In this case, there is a plane of parameters $1+4t_3+9t_6+16t_9 = 0$ on which the leading (non-constant) term in the small $n_\phi$ Hamiltonian expansion is $4$th-order in momentum, and the $SO(2)$ symmetry of the continuum Hamiltonian is reduced to the square lattice point group~\cite{Bauer22}. We refer to this as the \emph{quartic plane}. The $(t_3, t_6, t_9)=(-1/4,0,0)$ point on this plane corresponds to the ``zero-quadratic model" in Ref.~\onlinecite{Bauer22}. Similarly, on this plane, there is a line in parameter space $t_9=(1-15 t_6)/64$ on which the leading term is $6$th-order in momentum, which we call the \emph{hexic line}, and a point on this line $t_9=-1/56$ for which the leading term is $8$th-order in momentum, which we call the \emph{octic point}. Hence, the $(3,6,9)$ Hofstadter model allows us to explore a family of topological flat band models that do not have a Landau level continuum limit, and also comment on many-body behavior as we depart from this limit order-by-order.

\begin{figure}[h!]
	\includegraphics[width=\linewidth]{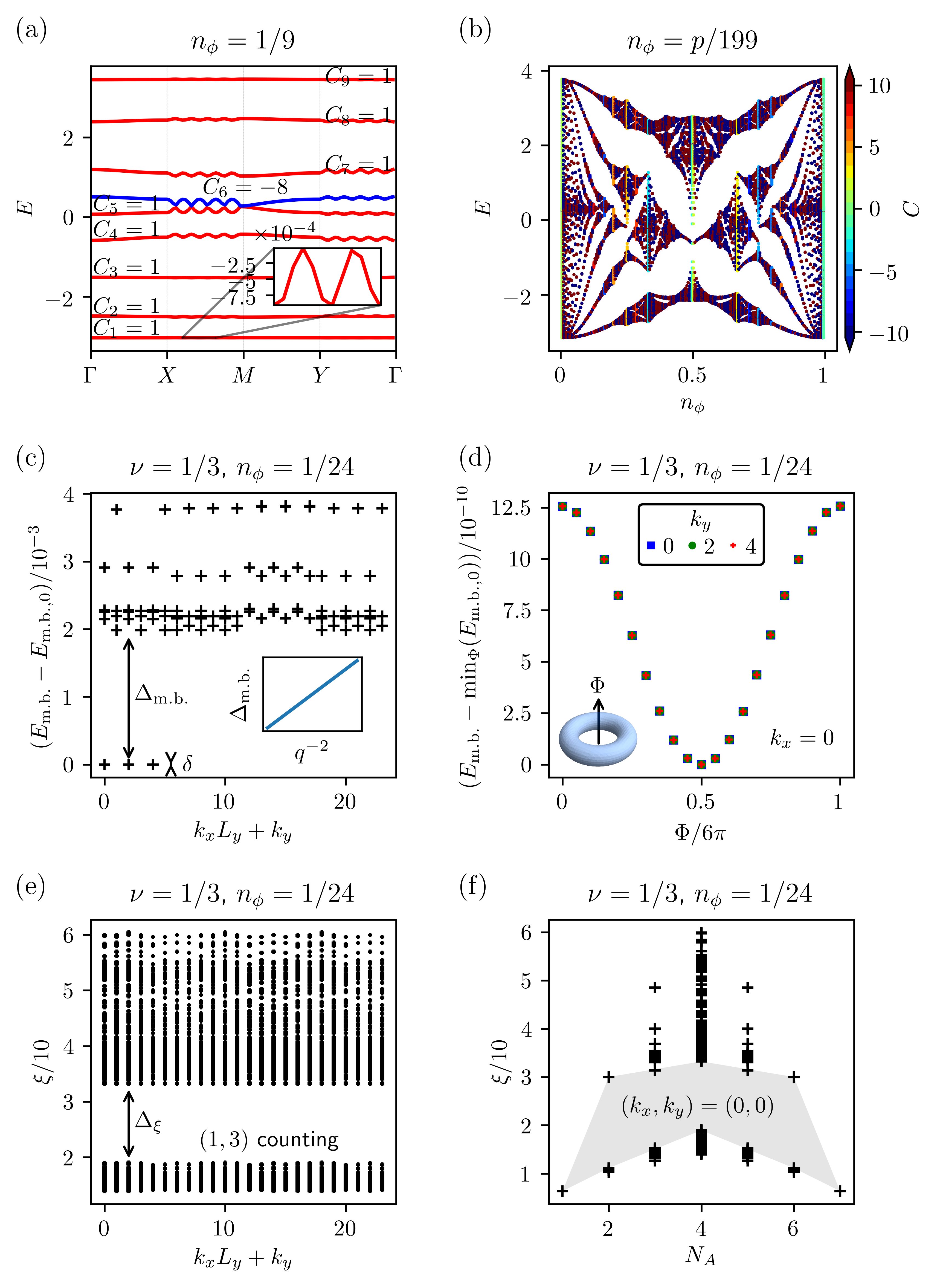}
	\caption{\label{fig:spmb} \textbf{Single- and many-body spectra on the hexic line.} (a)~Single-particle band structure of the $(3, 6, 9)$ Hofstadter model, defined as Eq.~\ref{eq:sp} with $t_1=1$, $n=\{1,3,6,9\}$, at the point $(t_3, t_6, t_9)=(-0.11, -0.15, 0.05)$ on the hexic line, with $n_\phi=1/9$. The Chern number $C$ of each band is labeled and the bands are colored according to its sign. Note that bands 5 and 6 are not touching. The fluctuations of the lowest band about its mean are shown inset. (b)~Butterfly plot showing the fractal single-particle energy spectrum as a function of $n_\phi$. The bands are colored from blue to red according to their Chern number, in the range $C \in [-10, 10]$. (c)~Many-body energy spectrum for the 8-particle fermionic Laughlin state filling $\nu=1/3$ of the lowest band at $n_\phi=1/24$, stabilized using nearest-neighbor interactions $V_{ij}=\delta_{\braket{ij}}$. The many-body gap $\Delta_\text{m.b.}$ and quasidegeneracy spread $\delta$ are labeled, and the gap scaling with magnetic unit cell area $q$ is shown inset. (d)~Spectral flow of the 3 quasidegenerate grounds states shown in (c), as a flux $\Phi$ is threaded through the handle of the torus. (e)~Particle entanglement spectrum corresponding to (c) obtained by tracing over half of the particles. The principal entanglement gap $\Delta_\xi$ is labeled and the number of states below the gap obeys the $(1, 3)$ counting rule~\cite{Regnault11}. (f)~Particle entanglement spectra corresponding to (c), in the $\mathbf{k}=\mathbf{0}$ momentum sector, obtained by tracing over $N_A$ particles. The entanglement gap is shaded gray.}
\end{figure}

In Fig.~\ref{fig:spmb}(a), we show the single-particle band structure of the $(3,6,9)$ Hofstadter model at the point $(t_3, t_6, t_9)=(-0.11, -0.15, 0.05)$ on the hexic line, with flux density $n_\phi=1/9$. The first Chern numbers of each band are labeled on the plot, and the bands are colored according to their sign. We note that the $(3,6,9)$ model has a topologically distinct band structure to the Hofstadter Hamiltonian, which would have the Chern numbers of the $5$th and $6$th bands interchanged to respect its symmetry. Nevertheless, we obtain a low-lying topological flat band with Chern number $C_1=1$, amenable to hosting the FQHE, as in the conventional Hofstadter case. In Fig.~\ref{fig:spmb}(b), we show the fractal energy spectrum of the $(3,6,9)$ model as a function of $n_\phi=p/199$ at the same point in parameter space, with Chern numbers in the range $C\in[-10, 10]$ colored from blue to red. Here we note that the center of the spectrum is shifted in the negative $E$ direction, when compared to the conventional Hofstadter butterfly, with subbands in the lower middle of the spectrum overlapping each other. This distortion reflects the slightly different band dispersion and Chern numbers resulting from the longer-range hoppings, as seen in Fig.~\ref{fig:spmb}(a). However, the broad properties of the spectrum are similar to the Hofstadter model, despite the fact that the $n_\phi\to 0,1$ lines no longer correspond to the Landau level limit. Overall, the single-particle properties of the $(3,6,9)$ model show that it has the potential to host FCIs, provided it is tuned to the appropriate system parameters. For a derivation of the single-particle band structure for the $(3,6,9)$ Hofstadter model, we refer the reader to Appendix~\ref{sec:sp_band_derivation}.  

\subsection{Many-body model}
\label{subsec:mb}

We introduce short-range particle interactions using minimal delta function terms in the Hamiltonian, such that the many-body system is described by
\begin{equation}
H = H_{(3,6,9)} + P_\text{LB} \sum_{ij} V_{ij} \normord{\rho_i \rho_j} P_\text{LB},
\end{equation} 
where $P_\text{LB}$ is the projector to the lowest band, $V_{ij}=\delta_{ij}(\delta_{\braket{ij}})$ is the contact(NN) interaction for bosons(fermions), and $\rho_i=c^\dagger_i c_i$ is the particle density operator at site $i$. It has been shown numerically for the Hofstadter model that neither the band projection nor the short-range interactions obstruct the stabilization of FCIs in the Jain series, owing to the favorable properties of the topological band structure~\cite{Andrews18}. Moreover, for simplicity, we restrict ourselves to lowest-lying topological flat bands with $C_1=1$ achieved at flux densities $n_\phi=1/q$, where the continuum limit corresponds to the Landau level limit in the conventional Hofstadter case.

In order to stabilize an FCI, it is crucial to define the correct lattice geometry and particle filling. We consider square system sizes, such that $N_x=N_y$, since this has been shown to be most stable for our numerical simulations~\cite{Scaffidi14}. Moreover, we focus on the primary states in the Jain series~\cite{Jain89, Sitko97}, the Laughlin states at filling $\nu=1/s$, with $s=2(3)$ for bosons(fermions), since these are the most robust and well-understood FQH states with the clearest signatures. For the bosonic Laughlin state, we consider MUCs of dimension $l_x\times l_y = m \times m$ with $L_x \times L_y = 4 \times 4$, whereas for the fermionic Laughlin state, we consider MUCs of dimension $l_x \times l_y = 3n\times 2n$ with $L_x \times L_y = 4 \times 6$. In both cases, we choose the integers $m, n$ to tune the flux density and we target the $\nu=N/N_\phi=1/2$ and $1/3$ FQH states for bosons and fermions, respectively, where $N_\phi=L_x\times L_y$ is the total number of flux quanta. We focus on Laughlin states with $N=8$ particles because it has been shown, through extensive exact diagonalization computations, that the many-body spectra of Laughlin FCIs in the Hofstadter model are already well-converged in the thermodynamic continuum limit at this system size (cf.~Figs.~2 and 6 of Ref.~\onlinecite{Andrews18}). Moreover, we would like to obtain a many-body example that is readily reproducible and comparable with previous studies~\cite{Jackson15, Bauer16, Bauer22}.

In Fig.~\ref{fig:spmb}(c), we show an example many-body energy spectrum of a fermionic Laughlin state stabilized at $n_\phi=1/24$. Here, the many-body energies $E_\text{m.b.}$ are offset with respect to the ground state energy $E_\text{m.b., 0}$ and plotted against the linearized momentum index $k_x L_y + k_y$~\footnote{The linearized momentum index is introduced simply to present the many-body spectrum on a two-dimensional plot. It is not a good quantum number.}. The ground-state degeneracy of a $\nu=r/s$ FQH state is given by $s^{g}$, where $g$ is the genus of the ground-state manifold~\cite{Bergholtz13}. On the torus ($g=1$), we observe a three-fold degenerate ground state clearly separated by a many-body gap $\Delta_\text{m.b.}$ to the higher-lying states. In practice, the degeneracy of these ground states is not exact, depending on physical factors, such as the proximity of an FCI to a phase transition~\cite{Grushin12, Andrews18}, as well as numerical factors, such as system size and numerical precision~\cite{Andrews18}. Therefore, we also define the quasidegeneracy spread of the ground states $\delta$ (although this spread is not visible on the scale of the plot). We can confirm the topological character of these ground states by threading a flux through the handle of the torus and verifying the spectral flow~\cite{Moller15, Regnault11}, as shown in Fig.~\ref{fig:spmb}(d). Finally, we note that due to the nature of the contact and NN interactions, the many-body gap scales $\Delta_\text{m.b.}\propto 1/q^{(2)}$ for bosons(fermions)~\cite{Jackson15, Bauer16, Andrews18}, as depicted in the inset of Fig.~\ref{fig:spmb}(c).

In Fig.~\ref{fig:spmb}(e), we show the particle entanglement spectrum (PES) corresponding to the FCI in Fig.~\ref{fig:spmb}(c). The spectrum is defined as the generalization of the particle-space Schmidt decomposition of a non-degenerate ground state $\ket{\Psi}=\sum_i e^{-\xi_i/2} \ket{\Psi^A_i}\otimes\ket{\Psi^B_i}$ to the degenerate case, such that $\{e^{-\xi/2}, \ket{\Psi^{A(B)}}\}$ now corresponds to the eigenbasis of the reduced density matrix $\rho_{A(B)}=\text{tr}_{B(A)}\rho$ with $\rho=\frac{1}{s}\sum_{i=1}^s\ket{\Psi_i}\bra{\Psi_i}$. By tracing over the positions of a subset of particles, it can be shown that the level counting corresponds exactly to the number of quasiholes in the system, and hence is a signature of FQH states. In FCIs, the relevant low-energy sector with precise quasihole counting is generally separated from the non-universal high-energy sector by a principal entanglement gap $\Delta_\xi$. In the case of the bosonic Laughlin state, we expect a counting according to the $(1,2)$ generalized Pauli exclusion principal, which for $N_A=4$ particles in $L_x\times L_y=4\times 4$ orbitals on a torus yields a total of $660$ quasihole states, whereas for the fermionic Laughlin state, we expect a $(1,3)$ counting, which for $N_A=4$ with $L_x\times L_y = 4\times 6$ yields a total of $2730$ states and is what we observe in Fig.~\ref{fig:spmb}(e)~\cite{Regnault11}. In Fig.~\ref{fig:spmb}(f), we show the PES in the $\mathbf{k}=\mathbf{0}$ momentum sector corresponding to Figs.~\ref{fig:spmb}(c,e) as we trace over $N_A$ particles. Here we can see how the number of states below the gap reduces for $N_A \neq 4$. For further details of the expected FCI quasihole counting, we refer the reader to Ref.~\onlinecite{Regnault11}.  

\section{Method}
\label{sec:method}

In this section, we outline our approach for probing the geometric stability hypothesis. In Sec.~\ref{subsec:sp_metrics}, we introduce the stability metrics in single-particle spectra, and analogously, in Sec.~\ref{subsec:mb_metrics}, we define our stability metrics in many-body spectra.

\subsection{Single-particle metrics}
\label{subsec:sp_metrics}

Since the numerical discovery of FCIs~\cite{Moller09, Neupert11, Sheng11, Tang11}, a number of single-particle metrics have been used to predict their stability. The most rudimentary of these is the flatness criterion $W\ll V \ll\Delta$, which predicts that, at sufficient interaction strength $V$, an increase in the gap-to-width ratio $\Delta/W$ will enhance FCI stability~\cite{Bergholtz13}. The intuition is that flat bands have a near-singular density of states, which together with a large $V$ and fractional filling maximizes particle interactions, while the large gap $\Delta$ prevents hopping to the next highest band. Although roughly true, several exceptions to this rule have been found, such as FCIs stabilized with $V\gg \Delta$~\cite{Andrews21_2}, and so this is not a precise indicator. Closely related to the gap-to-width ratio, the normalized Berry curvature fluctuations $\hat{\sigma}_\mathcal{B}=\sigma_\mathcal{B}/\mu_\mathcal{B}$ are also often used as an indicator of stability, where $\mu_\mathcal{B}$, $\sigma_\mathcal{B}$ correspond to the mean and standard deviation of the Berry curvature $\mathcal{B}$ defined over the Brillouin zone (BZ). The reasoning here is that Berry flatness is observed in Landau levels of the FQHE, which are known to exhibit robust fractional states~\cite{Roy14}.

Despite the frequent success of these two simple metrics, however, it was soon realized that band geometry plays an important role, in addition to band topology. Both of these properties are consolidated in the quantum geometric tensor (QGT), which underpins a diverse set of physical phenomena, from superfluidity to electrical conductance~\cite{Peotta15, Torma22, Julku21, Chan22, Mitscherling22, Mera22, Carollo20, Marzari12, Resta11, Albert16, Penner21}. In the context of Chern bands, the QGT is given as
\begin{equation}
\mathcal{R}_{ij}^\alpha (\mathbf{k}) = \left[ \partial_{k_i} \bra{\mathbf{k}, \alpha}\right] \mathcal{Q}_\alpha (\mathbf{k}) \left[ \partial_{k_j} \ket{\mathbf{k}, \alpha}\right],
\end{equation}
where $\alpha$ is the band index, $i,j$ are spatial indices, $\ket{\mathbf{k}, \alpha}$ is the Bloch eigenstate in band $\alpha$ with momentum $\mathbf{k}$, and $\mathcal{Q}_\alpha(\mathbf{k})=\mathbf{1}-\sum_{\beta\neq\alpha} \ket{\mathbf{k}, \beta}\bra{\mathbf{k}, \beta}$ is the orthogonal band projector. The real part of the QGT is given by the Fubini-Study metric $g^{\alpha}_{ij}(\mathbf{k})=\Re[\mathcal{R}^\alpha_{ij}]$, which corresponds to the distance between eigenstates on the Bloch sphere, whereas the imaginary part of the QGT is given by the Berry curvature $\mathcal{B}^\alpha(\mathbf{k})=-2 \Im [\mathcal{R}^\alpha_{xy}(\mathbf{k})]$. Crucially, since band geometry and topology are components of the same tensor, we can derive relations between them, namely
\begin{align}
\mathcal{D}(\mathbf{k})&=\text{det}\,g(\mathbf{k}) - \frac{1}{4}|\mathcal{B}(\mathbf{k})|^2 \geq 0, \\
\mathcal{T}(\mathbf{k})&=\text{tr}\,g(\mathbf{k}) - |\mathcal{B}(\mathbf{k})| \geq 0,
\end{align}
where we have dropped the band index $\alpha$ for simplicity, and we define $\mathcal{D}$ as the determinant inequality saturation measure (DISM) and $\mathcal{T}$ as the trace inequality saturation measure (TISM). It has been shown analytically that when the trace(determinant) inequality is saturated for a Chern band, the algebra of projected density operators is identical(isomorphic) to that in Landau levels~\cite{Roy14}. We note that since the trace inequality is the stronger condition of the two, the DISM is normally not considered. Based on this, it was conjectured that the closer the TISM is to zero, the more amenable a Chern band is to hosting FCI states~\cite{Jackson15}. This is known as the geometric stability hypothesis, which succinctly consolidates and extends all of the single-particle metrics that were used prior. Recently, more general numerical tests have been proposed to identify ideal Chern bands~\cite{Estienne23}, however it is not yet numerically shown that these provide a meaningful continuous metric away from the ideal point.

In light of this, we use the BZ-averaged TISM $\braket{\mathcal{T}}$ as our prominent single-particle stability metric. However, for reference we also compute: the BZ-averaged DISM $\braket{\mathcal{D}}$; the normalized Berry curvature fluctuations over the BZ $\hat{\sigma}_\mathcal{B}$, which is the leading-order contribution when mapping the algebra of projected density operators from Chern bands to Landau levels; the Fubini-Study metric fluctuations over the BZ $\sigma^2_g=\frac{1}{2}\sum_{ij}\sigma^2_{g_{ij}}$, which is the next-to-leading-order contribution; and the gap-to-width ratio $\Delta/W$. We define these quantities as in Ref.~\onlinecite{Jackson15}, for easy comparison.

\subsection{Many-body metrics}
\label{subsec:mb_metrics}

One of the main difficulties in testing the geometric stability hypothesis is that there is no unique way of defining the magnitude of FCI stability. Several methods have been employed in the literature, such as: the magnitude of the many-body gap $\Delta_\text{m.b.}$~\cite{Liu12, Andrews18, Jackson15, Bauer16, Bauer22, Moller15, Scaffidi12}, the magnitude of the many-body gap scaled by the quasidegeneracy spread $\Delta_\text{m.b.}/\delta$~\cite{Grushin12}, the magnitude of the principal entanglement gap $\Delta_\xi$~\cite{Andrews18, Bauer16, Scaffidi12, Dong23_2}, the critical interaction strength at which there is a phase transition $V_\text{crit}$~\cite{Grushin15, Andrews21_2}, and the extent to which a finite many-body gap $\Delta_\text{m.b.}$ persists in the large-$N$ limit~\cite{Liu12, Moller15, Andrews18}, to name a few. Although these metrics are often correlated, such as the magnitude of the many-body and entanglement gaps~\cite{Bauer16}, there are also cases when they are not. For example, we observe an infinite entanglement gap for the ground states of Laughlin and Read-Rezayi parent Hamiltonians, which becomes finite when we add a small perturbation to the Hamiltonian~\cite{Scaffidi12}, whereas the many-body gap will only be infinitesimally affected in this case. Hence, there is a need to unambiguously define a many-body metric in order to quantitatively compare FCI stability and comment on a stability hypothesis.

In our systems, we expect a quasidegeneracy spread of close to zero in all cases~\cite{Andrews18}, and so we cannot effectively use the $\Delta_\text{m.b.}/\delta$ metric. Based on empirical testing, we find that this measure works best as an FCI is tuned through a phase transition, where $\delta$ changes significantly. Nevertheless, we compute the quasidegeneracy spread in each case to verify that $\delta\approx 0$. Moreover, since phase transitions vary in nature and can be difficult to numerically characterize, it is not efficient to use the critical interaction strength $V_\text{crit}$ metric. Instead, we define FCI stability on fundamental terms, using the magnitude of gaps in the many-body and entanglement spectra, as shown in Figs.~\ref{fig:spmb}(c,e), which we spot check at larger $N$ to ensure that they are stable in the thermodynamic limit. We define the many-body gap as the gap between the largest $s$-fold quasidegenerate eigenenergy and the next highest state. Similarly, we define the principal entanglement gap as the gap between the largest of the quasihole entanglement eigenenergies and the next highest state. By using these metrics in unison, we can obtain a clearer picture of relative FCI stability with respect to the TISM, which will allow us to quantitatively evaluate the geometric stability hypothesis. Further details of the numerical method are described in Appendix~\ref{sec:num_method}.

\section{Results}
\label{sec:results}


In this section, we present a total of $735$ numerical exact diagonalization computations, solving for the many-body spectra of the $(3,6,9)$ Hofstadter Hamiltonian on a torus, and targeting the bosonic and fermionic Laughlin states in the lowest $C_1=1$ topological flat band at $n_\phi=1/q$. In all cases, we project the interaction Hamiltonian to the lowest band.

\begin{figure}
	\includegraphics[width=\linewidth]{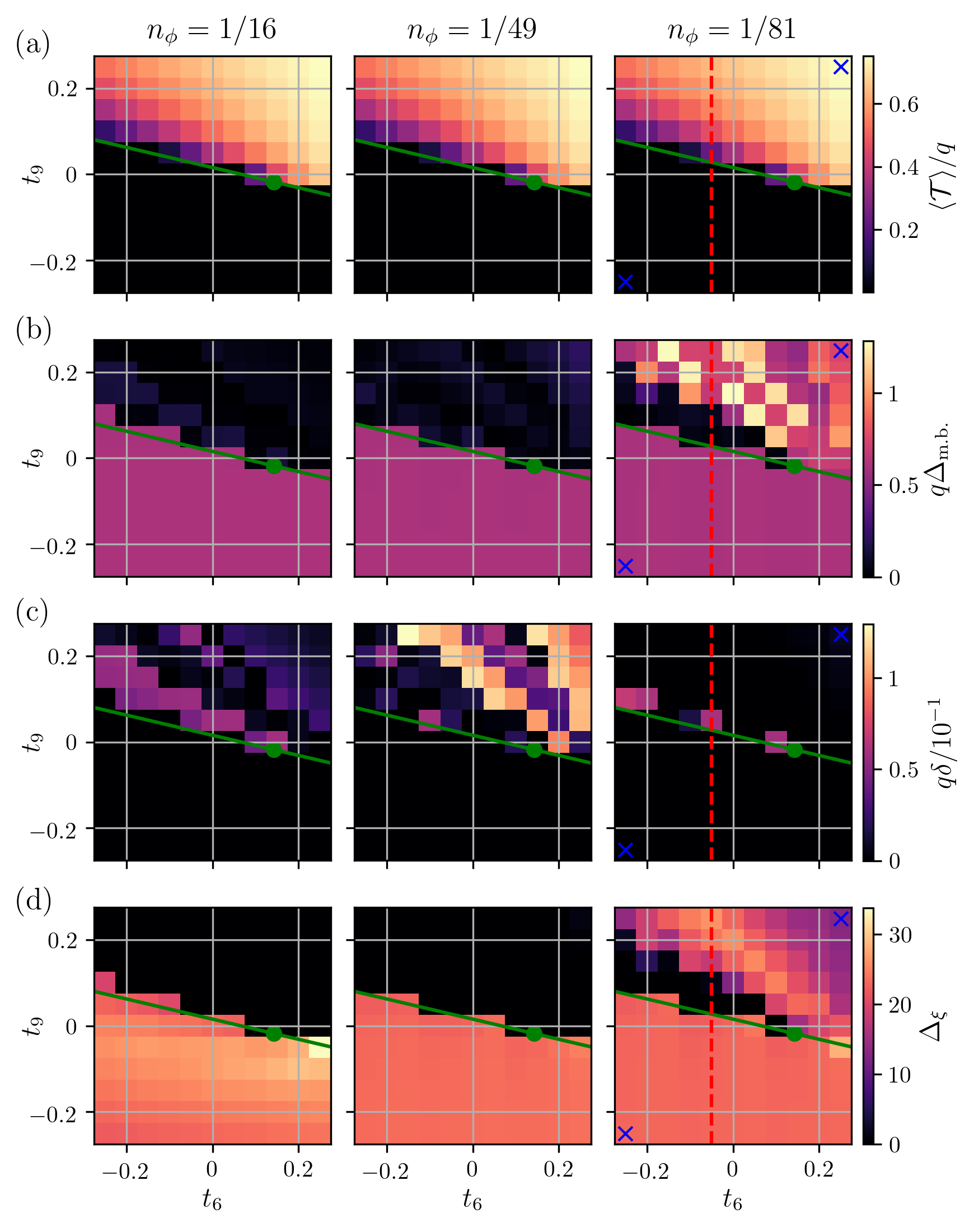}
	\caption{\label{fig:bosons} \textbf{Stability of the bosonic Laughlin state on the quartic plane.} (a)~BZ-averaged TISM $\braket{\mathcal{T}}$, scaled by the MUC area $q$, plotted in the quartic plane $1+4t_3+9t_6+16 t_9=0$, for $n_\phi=1/16, 1/49, 1/81$. The hexic line $t_9=(1-15t_6)/64$ and octic point $t_9=-1/56$ are overlaid in green. The points corresponding to Fig.~\ref{fig:case_study_bosons} are marked by blue crosses and the cross-section corresponding to Fig.~\ref{fig:laughlin_2d}(a) is marked by a red dashed line. (b)~Many-body gap $\Delta_\text{m.b.}$, scaled by the MUC area $q$, with parameters corresponding to (a). The results are shown for the 8-particle bosonic Laughlin state stabilized by the contact interaction $V_{ij}=\delta_{ij}$. (c)~Quasidegeneracy spread $\delta$, scaled by the MUC area $q$, corresponding to (b). (d)~Principal entanglement gap $\Delta_\xi$, corresponding to (b,c).}
\end{figure}

In Fig.~\ref{fig:bosons}, we show our comparison of single-particle and many-body stability metrics for the bosonic Laughlin state in the $(3,6,9)$ Hofstadter Hamiltonian on the quartic plane $1+4t_3+9t_6+16t_9=0$. We reiterate that unlike for the conventional Hofstadter model, for which the geometric stability hypothesis is known to hold, the single-particle bands do not converge to Landau levels as $n_\phi\to 0$ anywhere on this plane. In this figure, the top row (Fig.~\ref{fig:bosons}(a)) corresponds to our single-particle indicator, the BZ-averaged TISM, and the other three rows are derived from the many-body spectra. Figure~\ref{fig:bosons}(b) shows the many-body gap, Fig.~\ref{fig:bosons}(c) shows the quasidegeneracy spread, and Fig.~\ref{fig:bosons}(d) shows the principal entanglement gap. In Fig.~\ref{fig:bosons}(a), we can see that the scaled TISM $\braket{\mathcal{T}}/q$ converges in the continuum limit. The TISM is approximately zero for the region below the hexic line, where the quartic term has positive sign, and gradually increases as $t_6, t_9$ increase, in the region where the quartic term has negative sign. According to the geometric stability hypothesis, this would imply that FCIs are most stable below the hexic line and will either breakdown or decrease in stability as we move above the hexic line. From the many-body gap data in Fig.~\ref{fig:bosons}(b), this is what we observe at $n_\phi=1/16,1/49$. In these plots, we can see that there is a constant non-zero many-body gap below the hexic line, with a magnitude $q\Delta_\text{m.b.}\approx 0.6$ comparable to that of the Hofstadter model in the thermodynamic continuum limit~\cite{Andrews18, Bauer16}, which drops to approximately zero as we move above the line, signaling a breakdown of the FCI phase. The success of the geometric stability hypothesis here is perhaps unsurprising because at large $n_\phi$, the $(3,6,9)$ Hofstadter model behaves like the topological flat band models previously tested~\cite{Jackson15}. It is only as we approach the continuum limit $n_\phi\to 0$ that the effect of having no quadratic terms in momentum is accentuated. Indeed, as we move closer to the continuum at $n_\phi=1/81$, we witness anomalous behavior. In this case, we still observe a comparable non-zero many-body gap below the hexic line, however now there are also fluctuating non-zero many-body gaps above the hexic line. By taking vertical cross-sections of this plot, we can see that the magnitude of the many-body gap does not monotonically decrease as the TISM increases. Moreover, there are FCIs at $\braket{\mathcal{T}}\approx 0$ that have a smaller many-body gap than states with $\braket{\mathcal{T}}>0$. In order to investigate this observation, we plot the quasidegeneracy spread in Fig.~\ref{fig:bosons}(c). At small flux densities $n_\phi=1/16,1/49$, we can see a noisy distribution of $\delta$ values above the hexic line indicating unordered many-body spectra that are not in a topological phase. However, as we approach the continuum at $n_\phi=1/81$, we can see the quasidegeneracy spread reduces to zero in the region where we previously observed a non-zero many-body gap. Finally, we probe the phase diagram further by analyzing the entanglement spectra in Fig.~\ref{fig:bosons}(d). Corresponding to the plots above, we notice a clear non-zero principal entanglement gap above the correct $(1,2)$ counting in the region below the hexic line in all cases, which indicates the presence of an FCI phase. The gap magnitude $\Delta_{\xi} \approx 23$ converges as $n_\phi\to 0$ and is comparable to that for the Hofstadter model in the continuum limit (cf.~Fig.5(a) of Ref.~\onlinecite{Bauer16}). For $n_\phi=1/81$, however, we also observe a non-uniform non-zero gap region appearing above the hexic line. Moreover, as for the many-body gap data, there are points with $\braket{\mathcal{T}}>0$ that have a larger entanglement gap than in the $\braket{\mathcal{T}}\approx 0$ region. 

\begin{figure}
	\includegraphics[width=\linewidth]{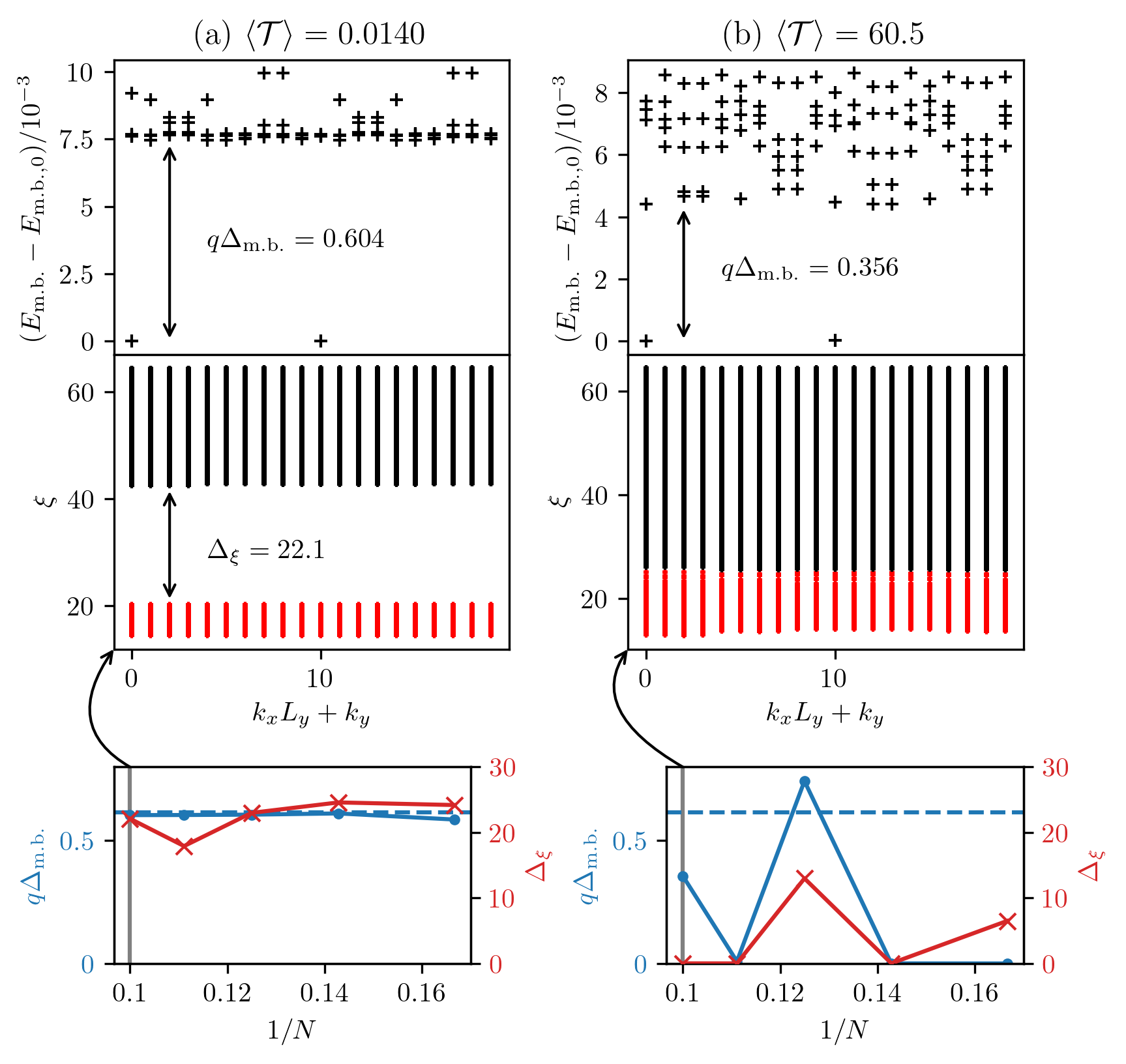}
	\caption{\label{fig:case_study_bosons} \textbf{Many-body spectra for candidate bosonic Laughlin states with small and large TISM.} Comparison of many-body energy and entanglement spectra for candidate 10-particle bosonic Laughlin FCIs at $n_\phi=1/81$, on the quartic plane, with (a)~$(t_6,t_9)=(-0.25, -0.25)$, and (b)~$(t_6,t_9)=(0.25,0.25)$, corresponding to the blue crosses in Fig.~\ref{fig:bosons}. The ground-state energies are (a)~$E_\text{m.b.,0}=-72.0$ and (b)~$E_\text{m.b.,0}=-74.6$. The entanglement energies included in the $(1, 2)$ counting are colored red. The bottom panels show the corresponding finite-size scaling of the many-body and entanglement gaps. The many-body gaps are marked by blue dots and the entanglement gaps are marked by red crosses. For reference, the scaled many-body gap of the Hofstadter model in the thermodynamic continuum limit is shown as a blue dashed line~\cite{Bauer16, Andrews18}. The lattice geometries are selected so that the total system is approximately square, with $|1-N_x / N_y|\leq 50\%$ in all cases. The PES are computed with $N_A=\lfloor N/2 \rfloor$.}
\end{figure}

At first glance, this may appear as though the geometric stability hypothesis has been violated. However, by performing a finite-size scaling, we find that the many-body and entanglement gaps in the region below the hexic line are stable in the thermodynamic limit, whereas the gaps in the region above the hexic line are unstable, as shown in Fig.~\ref{fig:case_study_bosons}. In particular, for the anomalous region, we observe that the entanglement gap closes as we increase particle number, which disqualifies it from an FCI phase~\cite{Jackson15}, whereas the entanglement gap in the lower region remains non-zero throughout the scaling and shows signs of convergence. Similarly, in Appendix~\ref{sec:add_res} we perform a flux density scaling and show that the anomalous region is also not robust in the continuum limit. This suggests that the geometric stability hypothesis holds in the thermodynamic and continuum limits, even for bands that are not continuously connected to Landau levels.

\begin{figure}
	\includegraphics[width=\linewidth]{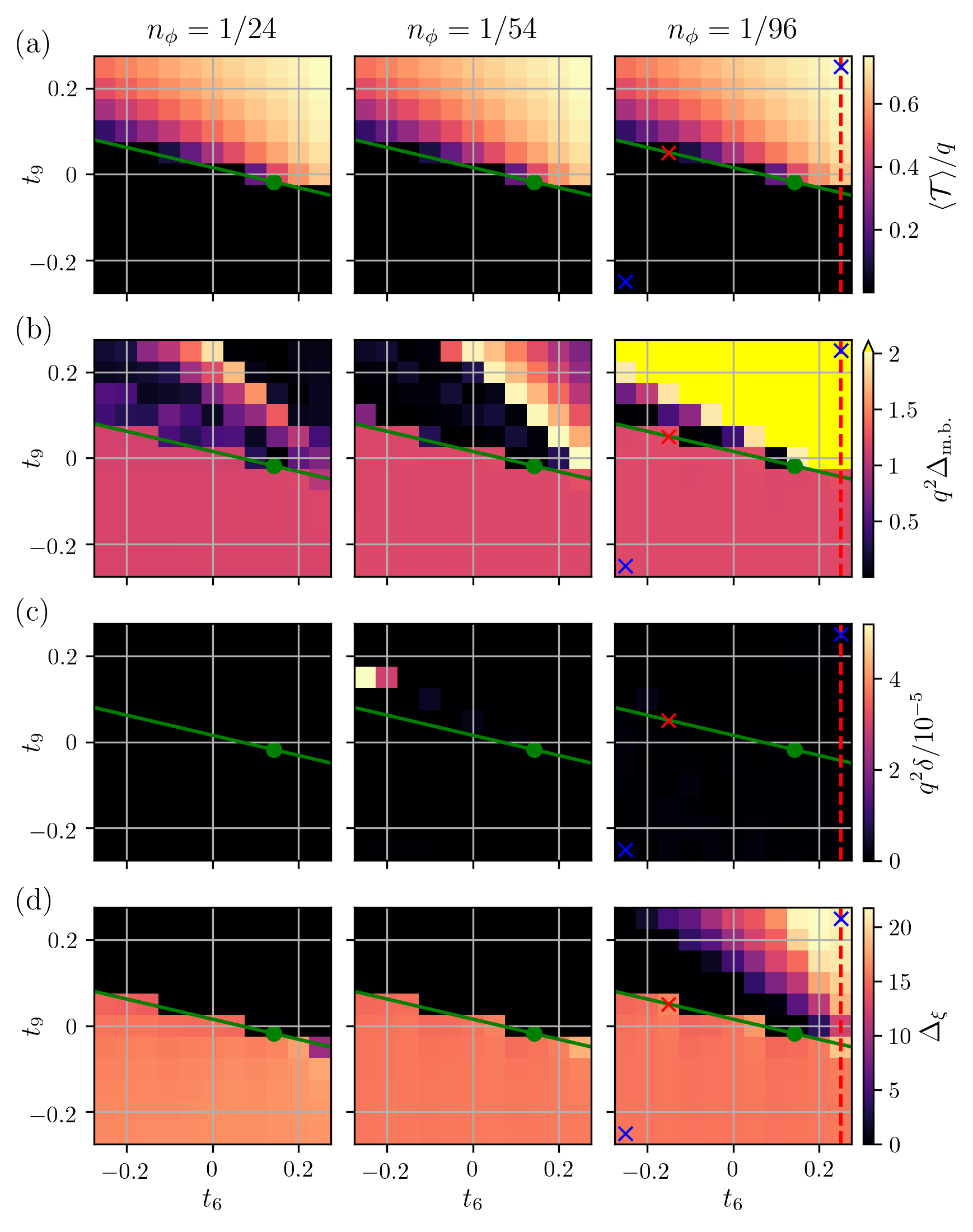}
	\caption{\label{fig:fermions} \textbf{Stability of the fermionic Laughlin state on the quartic plane.} (a)~BZ-averaged TISM $\braket{\mathcal{T}}$, scaled by the MUC area $q$, plotted in the quartic plane $1+4t_3+9t_6+16 t_9=0$, for $n_\phi=1/24, 1/54, 1/96$. The hexic line $t_9=(1-15t_6)/64$ and octic point $t_9=-1/56$ are overlaid in green. The point corresponding to Fig.~\ref{fig:spmb} is marked by a red cross, the points corresponding to Fig.~\ref{fig:case_study_fermions} are marked by blue crosses, and the cross-section corresponding to Fig.~\ref{fig:laughlin_2d}(b) is marked by a red dashed line. (b)~Many-body gap $\Delta_\text{m.b.}$, scaled by the MUC area $q$, with parameters corresponding to (a). The results are shown for the 8-particle fermionic Laughlin state stabilized by the nearest-neighbor interaction $V_{ij}=\delta_{\braket{ij}}$. (c)~Quasidegeneracy spread $\delta$, scaled by the MUC area $q$, corresponding to (b). (d)~Principal entanglement gap $\Delta_\xi$, corresponding to (b,c).}
\end{figure}

In Fig.~\ref{fig:fermions}, we show the plot for the fermionic Laughlin state corresponding to Fig.~\ref{fig:bosons}. As before, we see in Fig.~\ref{fig:fermions}(a) that the scaled TISM $\braket{\mathcal{T}}/q$ converges in the continuum limit. From the many-body gap data in Fig.~\ref{fig:fermions}(b), we notice a similar trend to the bosonic Laughlin state in Fig.~\ref{fig:bosons}(b). At all values of flux density, we observe a constant non-zero many-body gap below the hexic line, with a similar magnitude $q^2\Delta_\text{m.b.}\approx1.2$ to that for the Hofstadter model in the thermodynamic continuum limit~\footnote{We note that the many-body gap magnitudes for the fermionic Laughlin state in Ref.~\onlinecite{Andrews18} differ by a factor of two compared to our study and Ref.~\onlinecite{Bauer16}. Moreover, the Schmidt values $\lambda_i$ are defined as $e^{-\xi_i}$ in Ref.~\onlinecite{Andrews18}, whereas they are defined as $e^{-\xi_i/2}$ in our study and Ref.~\onlinecite{Bauer16}. Hence, the principal entanglement gaps also differ by a factor of two.} (cf.~Fig.4(b) of Ref.~\onlinecite{Bauer16}). At intermediate values of flux density $n_\phi=1/24, 1/54$, we can see that this non-zero many-body gap diminishes above the hexic line, in agreement with the geometric stability hypothesis, apart from a few sporadic points, which we do not categorize as FCIs due to an incorrect quasihole counting. However, as we approach the continuum limit at $n_\phi=1/96$, we can see that there is a region above the hexic line that has a non-zero many-body gap, which is approximately an order of magnitude larger than the gap observed for $\braket{\mathcal{T}}\approx 0 $~\footnote{This anomalous region above the hexic line is also non-uniform, although not visible on the scale of the plot.}. Data for the quasidegeneracy spread are less insightful here, since $\delta\approx 0$ in almost all cases, however they do not eliminate the possibility of an FCI phase. Finally, in Fig.~\ref{fig:fermions}(d), we notice an analogous trend to that seen in Fig.~\ref{fig:bosons}. At flux densities $n_\phi=1/24,1/54$, we observe the correct $(1,3)$ counting below the hexic line with a clear non-zero entanglement gap that converges as $n_\phi\to 0$, whereas above the hexic line the gap goes to zero, indicating a breakdown of the FCI phase. This shows that the points of non-zero many-body gap above the hexic line, observed in the left two panels of Fig.~\ref{fig:fermions}(b), do not correspond to FCIs. As we approach the continuum at $n_\phi=1/96$, we can again see the constant region of non-zero entanglement gap $\Delta_{\xi}\approx 16$ below the hexic line, with a magnitude comparable to that for the Hofstadter model in the continuum limit (cf.~Fig.5(b) of Ref.~\onlinecite{Bauer16}). Moreover, there is now also an anomalous non-uniform region of non-zero gap above the hexic line, with a significantly larger entanglement gap. 

\begin{figure}
	\includegraphics[width=\linewidth]{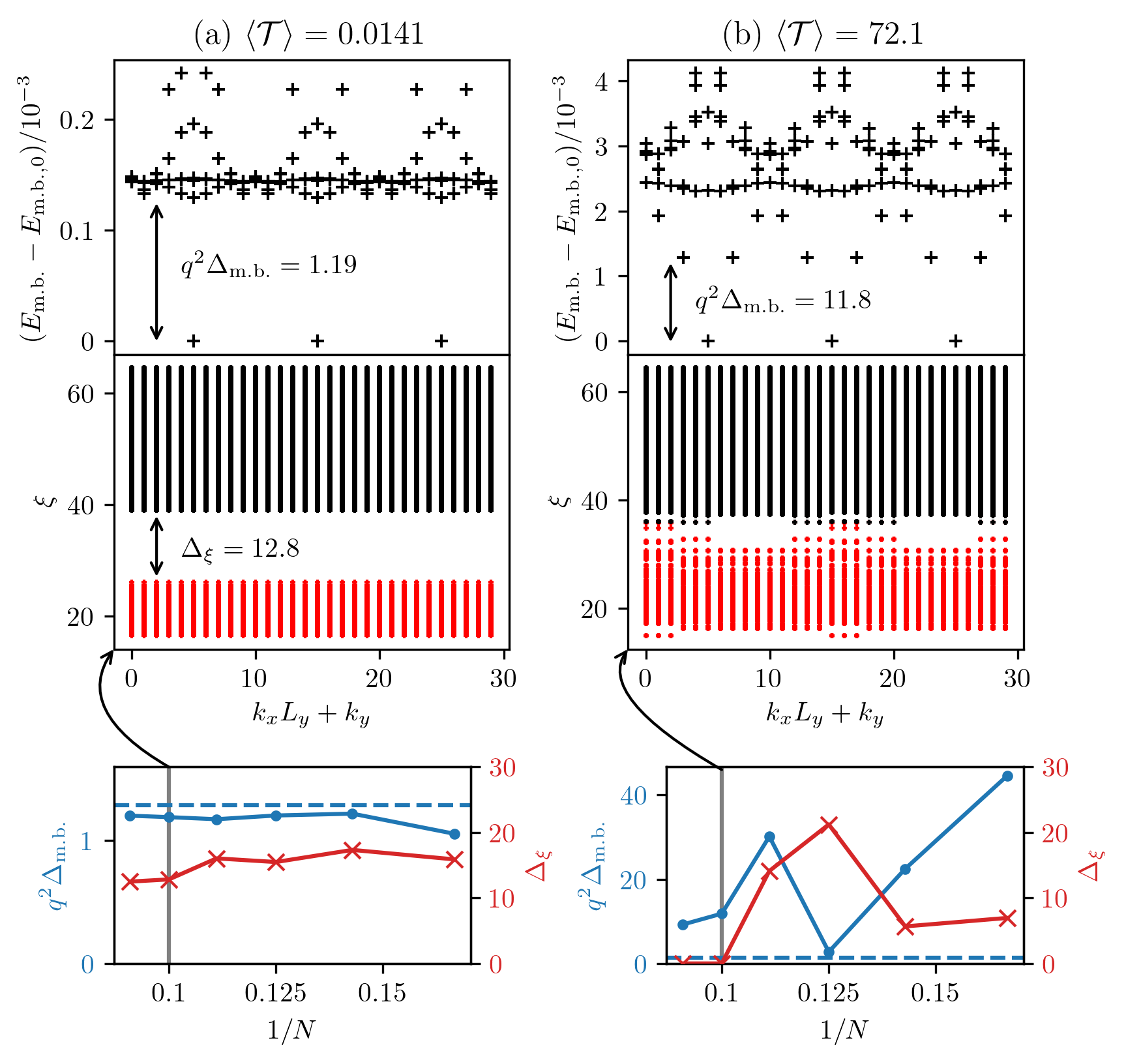}
	\caption{\label{fig:case_study_fermions} \textbf{Many-body spectra for candidate fermionic Laughlin states with small and large TISM.} Comparison of many-body energy and entanglement spectra for candidate 10-particle fermionic Laughlin FCIs at $n_\phi=1/96$, on the quartic plane, with (a)~$(t_6,t_9)=(-0.25, -0.25)$, and (b)~$(t_6,t_9)=(0.25,0.25)$, corresponding to the blue crosses in Fig.~\ref{fig:fermions}. The ground-state energies are (a)~$E_\text{m.b.,0}=-72.1$ and (b)~$E_\text{m.b.,0}=-75.8$. The entanglement energies included in the $(1, 3)$ counting are colored red. The bottom panels show the corresponding finite-size scaling of the many-body and entanglement gaps. The many-body gaps are marked by blue dots and the entanglement gaps are marked by red crosses. For reference, the scaled many-body gap of the Hofstadter model in the thermodynamic continuum limit is shown as a blue dashed line~\cite{Bauer16, Andrews18}. The lattice geometries are selected so that the total system is approximately square, with $|1-N_x / N_y|\leq 27\%$ in all cases. The PES are computed with $N_A=\lfloor N/2 \rfloor$.}
\end{figure}

As before, this may appear as though the geometric stability hypothesis has been violated at first glance. However, by performing a finite-size scaling, we can see that the many-body and entanglement gaps are stable in the region below the hexic line, whereas they fluctuate wildly in the region above the hexic line, as shown in Fig.~\ref{fig:case_study_fermions}. In particular, the entanglement gap closes in the anomalous region as we increase particle number, which disqualifies it from an FCI phase~\cite{Jackson15}, whereas the entanglement gap remains non-zero and shows signs of convergence in the lower region. In addition, the flux density scaling in Appendix~\ref{sec:add_res} shows that the anomalous region is also not robust in the continuum limit. We emphasize that the system sizes and flux densities chosen in Figs.~\ref{fig:bosons} and~\ref{fig:fermions} correspond to those used in previous numerical studies of the geometric stability hypothesis~\cite{Jackson15, Bauer16, Bauer22}, which highlights the additional care required when analyzing models with a non-Landau level continuum limit. Ultimately, these data again support the geometric stability hypothesis in the thermodynamic and continuum limits, even for bands not continuously connected to Landau levels.

\begin{figure}
	\includegraphics[width=\linewidth]{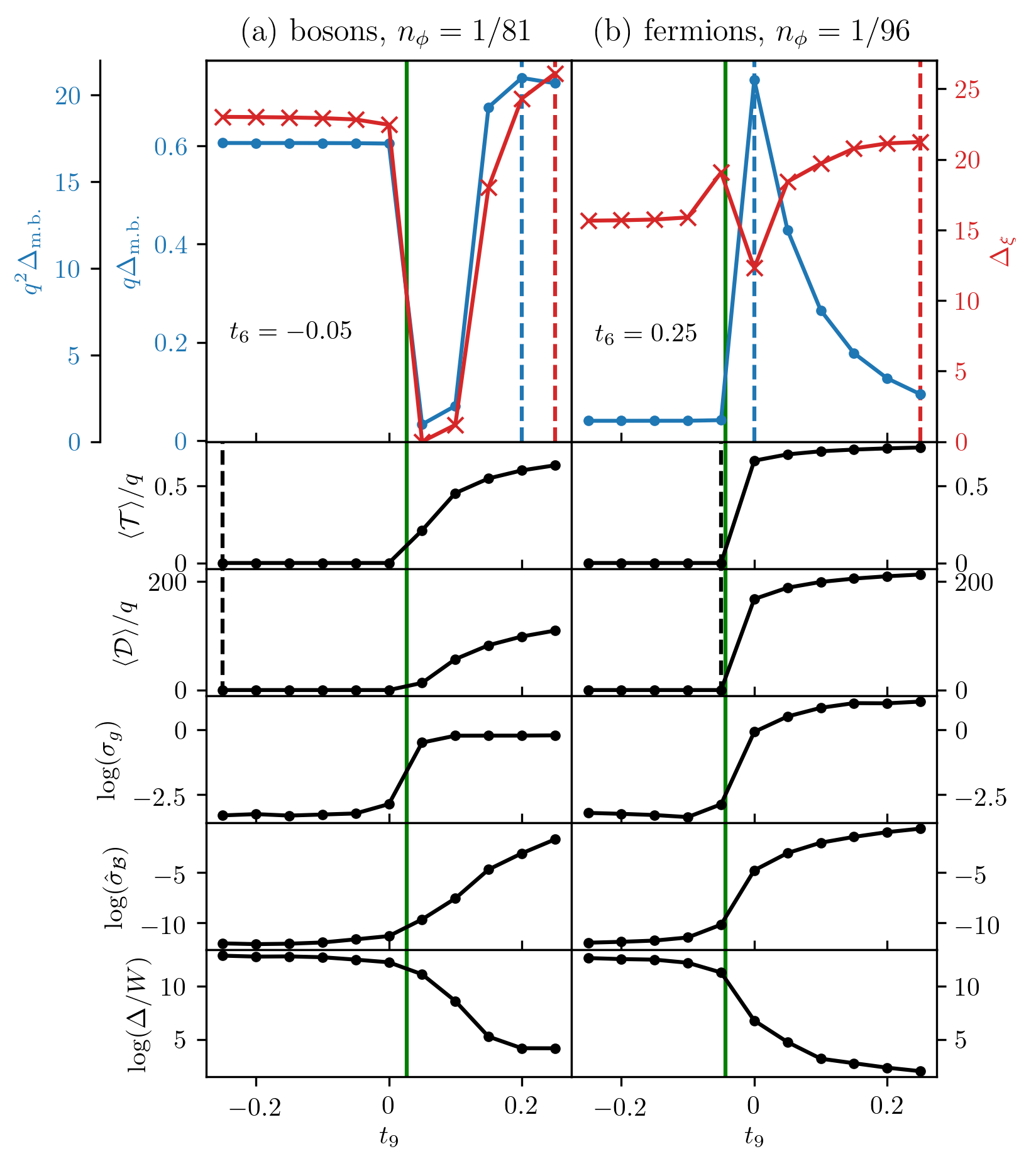}
	\caption{\label{fig:laughlin_2d} \textbf{Stability of Laughlin states on cross-sections of the quartic plane.} Comparison of cross-sections at (a)~$t_6=-0.05$ for the bosonic Laughlin state at $n_\phi=1/81$, and (b)~$t_6=0.25$ for the fermionic Laughlin state at $n_\phi=1/96$, corresponding to the red dashed lines in Figs.~\ref{fig:bosons} and~\ref{fig:fermions}. The first panels show the scaled many-body gap, $q^{(2)}\Delta_\text{m.b.}$ for bosons(fermions), and the principal entanglement gap $\Delta_\xi$, as a function of $t_9$. The many-body gaps are marked by blue dots and the entanglement gaps are marked by red crosses. The maximum values of the gaps are marked with dashed lines, and the hexic line is marked with a solid green line, as in Figs.~\ref{fig:bosons} and~\ref{fig:fermions}. The subsequent panels show the corresponding single-particle metrics in black. The second, third, and fourth panels show the scaling of the BZ-averaged TISM $\braket{\mathcal{T}}$, the BZ-averaged DISM $\braket{\mathcal{D}}$, and fluctuations of the Fubini-Study metric $\sigma_g$, respectively. The minima of the TISM and DISM are marked with a dashed line. Finally, the fifth and sixth panels show the scaling of the normalized Berry curvature fluctuations $\hat{\sigma}_\mathcal{B}$ and the gap-to-width ratio $\Delta / W$. The logarithmic scale on the $y$-axes is base $10$.} 
\end{figure}

To explicitly demonstrate the spurious violations of the geometric stability hypothesis at small system sizes, we focus on the cross-sections marked by the red dashed lines in Figs.~\ref{fig:bosons} and~\ref{fig:fermions}. We start with the cross-section for the bosonic Laughlin state from Fig.~\ref{fig:bosons} at $n_\phi=1/81$, which is plotted in Fig.~\ref{fig:laughlin_2d}(a). According to the geometric stability hypothesis, we would expect FCI stability to monotonically increase as $\braket{\mathcal{T}}\to 0$. Instead, there appears to be a phase transition, with a trivial phase observed at $t_9=0.05$~\footnote{The small but finite entanglement gap at this point reflects the energy scale of the spectrum and is not a topological signature.}, suggesting that a monotonic decrease in the TISM does not lead to a monotonic increase in FCI stability. Moreover, in this case we notice that both of our FCI stability metrics, i.e.~the many-body gap and the entanglement gap, are correlated, and they both seem to indicate that states at large $\braket{\mathcal{T}}\approx 50$ are more stable than at $\braket{\mathcal{T}}\approx0$. For comparison, we also plot the other single-particle stability metrics discussed in Sec.~\ref{subsec:sp_metrics}. Here we can see that the DISM is strongly correlated with the TISM, and the gap-to-width ratio is inversely correlated with the Berry fluctuations, as expected. We note that in this example, the primitive single-particle stability metrics, such as the gap-to-width ratio and Berry fluctuations accord with the band geometric measures, such as the TISM, DISM, and Fubini-Study fluctuations, which is not always the case~\cite{Ledwith22}. For all metrics, we notice a clear transition about the hexic line, where the quartic term changes sign. Similarly, we can examine the cross-section for the fermionic Laughlin state from Fig.~\ref{fig:fermions} at $n_\phi=1/96$, which is plotted in Fig.~\ref{fig:laughlin_2d}(b). As in Fig.~\ref{fig:laughlin_2d}(a), we can see that FCI stability does not appear to monotonically increase as $\braket{\mathcal{T}}\to 0$, regardless of which many-body stability metric is considered. Interestingly, on this occasion we can see that the many-body and entanglement gaps are inversely correlated above the hexic line. This reflects the rapid reduction of the many-body gap as $t_6, t_9\gg 0$ are increased, which is obscured by the color scale in the right panel of Fig.~\ref{fig:fermions}(b). Nevertheless, there are points with $\braket{\mathcal{T}}\approx50$ for which both the many-body and entanglement gaps are larger than at $\braket{\mathcal{T}}\approx 0$. Again, we can see that all single-particle metrics accord with each other, showing a distinct transition about the hexic line. Comparing FCIs at quartic, hexic, and octic points with fixed flux density in Figs.~\ref{fig:bosons} and~\ref{fig:fermions}, we do not notice a significant change in the many-body and entanglement gaps, although these Hamiltonian properties do have an overall effect on the phase diagram.

\section{Discussion \& Conclusions}
\label{sec:conc}

In this paper, we have investigated the scope of the geometric stability hypothesis by analyzing the stability of a family of FCIs that do not have a Landau level continuum limit. In Sec.~\ref{subsec:sp}, we started by explaining why the $(3,6,9)$ Hofstadter model in the quartic plane has a non-Landau level continuum limit. In Sec.~\ref{subsec:mb}, we then showed how the flat bands of this model are amenable to hosting FCIs given an appropriate lattice geometry, filling fraction, and interaction Hamiltonian. Subsequently, in Sec.~\ref{sec:method}, we outlined our method for testing the geometric stability hypothesis. In Sec.~\ref{subsec:sp_metrics}, we explained why the TISM is the prominent single-particle stability metric, and in Sec.~\ref{subsec:mb_metrics}, we justified our choice of many-body and entanglement gaps as quantifiers of FCI stability. Lastly, in Sec.~\ref{sec:results}, we analyzed the stability of Laughlin states in our model and demonstrated that the geometric stability hypothesis also applies to non-Landau flat bands, albeit often requiring larger system sizes to converge for these configurations.

In Sec.~\ref{sec:results}, we chose system sizes and flux densities that correspond to previous numerical studies of the geometric stability hypothesis for topological flat band models with a Landau level continuum limit~\cite{Jackson15, Bauer16, Bauer22}. In doing so, we are able to directly compare numerical aspects of the simulations. We find that, even though $N=8$ Laughlin FCIs are well-converged in the thermodynamic limit for the conventional Hofstadter model~\cite{Andrews18}, this is not the case for the $(3,6,9)$ model. In Figs.~\ref{fig:bosons} and~\ref{fig:fermions}, we notice anomalous regions above the hexic line that are not robust in the $N\to\infty$ limit, and in Appendix~\ref{sec:add_res}, we show that these regions also diminish as we continue to take the $n_\phi\to0$ limit. Since these anomalous regions are fluctuating, highly-sensitive to numerical precision, and not robust in the thermodynamic or continuum limits, we attribute this to a numerical instability and not a physical phase transition. Moreover, since the relative magnetic length in our simulations is constant, the most likely cause of this transient numerical instability is the relative dominance of the fourth-order momentum term in the Hamiltonian.       

As previously mentioned, the definition of FCI stability is ambiguous, which makes quantitatively testing the geometric stability hypothesis problematic. We decided to use the many-body and entanglement gaps as quantifiers because they are direct properties of the many-body Hamiltonian, despite the fact that they are not always correlated. Other methods of quantifying FCI stability, such as the range of interaction strengths over which an FCI is stabilized~\cite{Andrews21_2, Grushin15}, may lead to different results. Another factor that is important to consider is the form of the interaction term. For example, an FCI at $\braket{\mathcal{T}}=0$ will not be more stable than an FCI at $\braket{\mathcal{T}}>0$ if its interaction terms are significantly less optimal. In our investigation, we have compared FCIs with identical contact / NN interactions to remove this variable. We also reiterate that, despite the plethora of recent extensions and generalizations~\cite{Simon20, Ledwith22, Estienne23, Fujimoto24, Abouelkomsan23}, the original geometric stability hypothesis holds up well for models that are not continuously connected to Landau levels in our study, modulo misleading breakdowns that are of numerical origin. 

We have presented results highlighting the limitations of testing the geometric stability hypothesis using finite-size numerics. In future work, it would be interesting to develop and test the efficacy of continuous single-particle stability metrics in more general frameworks~\cite{Simon20, Ledwith22, Estienne23, Fujimoto24}, analogous to the TISM, since most current research is focused on criteria for the ideal Chern band itself~\cite{Mera21, Estienne23}. Moreover, although we have presented a simple example of a model that does not have a Landau level continuum limit, this is not the only example. The conventional Hofstadter model famously has an effective continuum for bands of higher Chern number in the limit $n_\phi\to1/|C|$~\cite{Andrews18}. It would therefore be interesting to check whether there is also a finite-size breakdown in the geometric stability hypothesis here, although this would be more challenging, since it would require systematically stabilizing a large set of higher-$|C|$ FCIs~\cite{Dong23, Sterdyniak13, Liu12, Andrews18, Moller15}. We hope that showcasing the scope of the geometric stability hypothesis motivates further efforts toward its generalization, along with a reliable and inexpensive single-particle metric, that can be used to universally compare the relative stability of FCIs.

\begin{acknowledgments}
	We thank Daniel Parker, Glenn Wagner, Gunnar M\"oller, Johannes Mitscherling, Jie Wang, and David Bauer for useful discussions. In particular, we thank Daniel Parker and Glenn Wagner for helpful comments on the manuscript. The single-particle computations were performed using HofstadterTools~\cite{HofstadterTools} and the many-body computations were performed using DiagHam. B.A.~acknowledges support from the Swiss National Science Foundation under grant no.~P500PT\_203168, and B.A.~and R.R.~acknowledge support from the University of California Laboratory Fees Research Program funded by the UC Office of the President (UCOP), grant no.~LFR-20-653926. B.A.~and M.R.~are grateful for the funding received from the UC Berkeley Physics Innovators Initiative (Pi\textsuperscript{2}) program. M.Z.~and B.A.~were funded by the U.S. Department of Energy, Office of Science, Office of Basic Energy Sciences, Materials Sciences and Engineering Division under contract no.~DE-AC02-05-CH11231 (Theory of Materials program KC2301).
\end{acknowledgments}

\appendix

\section{Single-particle band structure of the $(3,6,9)$ Hofstadter model}
\label{sec:sp_band_derivation}

In this section, we consider the $(3,6,9)$ Hofstadter model, defined as Eq.~\ref{eq:sp} with $n=\{1, 3, 6, 9\}$ and $t_1=1$.

As mentioned in the main text, the Peierls phase is defined as $\theta_{ij}=(2\pi/\phi_0)\int_i^j \mathbf{A}\cdot\mathrm{d}\mathbf{l}$, where $\mathbf{A}$ is the vector potential and $\mathrm{d}\mathbf{l}$ is an infinitesimal line element along the path from $i=(X_i, Y_i)$ to $j=(X_j, Y_j)$. Performing this line integral in Landau gauge $\mathbf{A}=Bx\hat{\mathbf{e}}_y$ yields the general formula
\begin{equation}
\theta_{ij} = 2 \pi n_\phi (Y_j-Y_i)\left( X_i + \frac{X_j - X_i}{2}\right),
\end{equation}
where the flux density $n_\phi=Ba^2/\phi_0=p/q$ with coprime integers $(p,q)$, $\phi_0$ is the flux quantum, and $X, Y$ are measured in units of $a$. Note that, in this choice of gauge, the Peierls phase depends on relative $y$-coordinates but absolute $x$-coordinates. Using this formula, we find that the only hoppings in the $(3,6,9)$ Hofstadter model with a non-zero Peierls phase are given as
\begin{equation}
\theta_{ij}^{(m,n)} = \begin{cases}
2\pi n_\phi m, & j=(m, n+1), \\
4\pi n_\phi m, & j=(m, n+2), \\
6\pi n_\phi m, & j=(m, n+3), \\
8\pi n_\phi m, & j=(m, n+4),
\end{cases}
\end{equation}
where we have defined $i = (m,n)$.

Taking the plane wave ansatz $\Psi_{m,n}=e^{\mathrm{i}k_x m a}e^{\mathrm{i}k_y n a}\psi_m$, and substituting this into the time-independent Schr\"odinger equation $H\Psi_{m,n}=E\Psi_{m,n}$, leaves us with 
\begin{equation}
\begin{split}
E\psi_m = F^* \psi_{m-4} + D^* \psi_{m-3} + C^* \psi_{m-2} \\ + B^* \psi_{m-1} + A_m \psi_m + B \psi_{m+1} \\ + C \psi_{m+2} + D \psi_{m+3} + F \psi_{m+4},
\end{split}
\end{equation}
where
\begin{align}
A_m &= -2\sum_{\tau=1}^4 t'_\tau \cos(2\tau \pi n_\phi m +\tau k_y a), \\
B &= -t'_1 e^{\mathrm{i}k_x a}, \\
C &= -t'_2 e^{\mathrm{i}2 k_x a}, \\
D &= -t'_3 e^{\mathrm{i}3 k_x a}, \\
F &= -t'_4 e^{\mathrm{i}4 k_x a},
\end{align}
and we have skipped the letter $E$ to avoid confusion with the eigenenergies. Consequently, the full eigenvalue problem $H \boldsymbol{\psi} = E \boldsymbol{\psi}$, where $\boldsymbol{\psi} = (\psi_1, \psi_2, \dots, \psi_{q})^\intercal$, is described by the $q\times q$ Hamiltonian matrix
\begin{widetext}
\begin{equation}
\mathbf{H} =
\begin{pmatrix}
A_1 & B & C & D & F & \dots & 0 & F^* & D^* & C^* & B^* \\
B^* & A_2 & B & C & D & \dots & 0 & 0 & F^* & D^* & C^* \\
C^* & B^* & A_3 & B & C & \dots & 0 & 0 & 0 & F^* & D^* \\
D^* & C^* & B^* & A_4 & B & \dots & 0 & 0 & 0 & 0 & F^* \\
F^* & D^* & C^* & B^* & A_5 & \dots & 0 & 0 & 0 & 0 & 0 \\
\vdots & \vdots & \vdots & \vdots & \vdots & \ddots & \vdots & \vdots & \vdots & \vdots & \vdots \\
0 & 0 & 0 & 0 & 0 & \dots & A_{q-4} & B & C & D & F \\
F & 0 & 0 & 0 & 0 & \dots & B^* & A_{q-3} & B & C & D \\
D & F & 0 & 0 & 0 & \dots & C^* & B^* & A_{q-2} & B & C \\
C & D & F & 0 & 0 & \dots & D^* & C^* & B^* & A_{q-1} & B \\
B & C & D & F & 0 & \dots & F^* & D^* & C^* & B^* & A_q \\
\end{pmatrix}.
\end{equation}
\end{widetext}
The eigenvalues of the above Hamiltonian yield the $q$-band single-particle energy spectrum for the $(3,6,9)$ Hofstadter model with $n_\phi=p/q$, shown in Figs.~\ref{fig:spmb}(a,b).

\section{Details of the numerical method}
\label{sec:num_method}

We compute the many-body spectra of the $(3,6,9)$ Hofstadter model using the Lanczos algorithm implemented in the DiagHam package. In most cases, we use a Lanczos precision of $\Delta E_\text{m.b.} = 10^{-10}$ and compute the lowest five eigenstates in each momentum sector. However, there are a few cases, where we need to either adjust the precision or compute a larger number of eigenstates in order to ensure that the low-lying states are properly converged, such as for the $N=9$ point in Fig.~\ref{fig:case_study_fermions}(a). This can have a significant effect on the energy and entanglement spectra, with the latter being more sensitive.

For the systematic computations on the quartic plane, shown in Figs.~\ref{fig:bosons},~\ref{fig:fermions},~\ref{fig:laughlin_2d}, and~\ref{fig:bosons_ext}, we use a square total system size, as described in the main text. However, when performing the finite-size scaling in Figs.~\ref{fig:case_study_bosons} and~\ref{fig:case_study_fermions}, or the flux density scaling for fermions in Fig.~\ref{fig:fermions_ext}, this is not always possible for the given particle numbers and flux densities, and so we use approximately square systems. We define the squareness deviation parameter as
\begin{equation}
\epsilon = \left| 1 - \frac{N_x}{N_y} \right|.  
\end{equation}
For each system size, we search for lattice geometries that yield the correct filling fraction $L_x L_y / N =s$ and flux density $l_x l_y = q$, while minimizing $\epsilon$. For the bosonic Laughlin state in Fig.~\ref{fig:case_study_bosons}, lattice geometries that satisfy this for $1\leq L_x,l_x,L_y,l_y < 100$ are given in Table~\ref{tab:lat_geom}(a). For the fermionic Laughlin state in Fig.~\ref{fig:case_study_fermions}, corresponding lattice geometries are given in Table~\ref{tab:lat_geom}(b).

\begin{table}
	\begin{center}
		(a) bosons
	\end{center}
	\vspace{-0.5em}
	\begin{ruledtabular}
		\begin{tabular}{c c c c c c}
			$N$ & $l_x$ & $l_y$ & $L_x$ & $L_y$ & $\epsilon$ \\
			\hline
			6 & 9 & 9 & 3 & 4 & 0.25 \\
			7 & 27 & 3 & 1 & 14 & 0.36 \\
			8 & 9 & 9 & 4 & 4 & 0 \\
			9 & 9 & 9 & 3 & 6 & 0.5 \\
			10 & 9 & 9 & 4 & 5 & 0.20 \\
		\end{tabular}
	\end{ruledtabular}
	
	\begin{center}
		(b) fermions
	\end{center}
	\vspace{-0.5em}
	\begin{ruledtabular}
		\begin{tabular}{c c c c c c}
			$N$ & $l_x$ & $l_y$ & $L_x$ & $L_y$ & $\epsilon$ \\
			\hline
			6 & 12 & 8 & 3 & 6 & 0.25 \\
			7 & 16 & 6 & 3 & 7 & 0.14 \\
			8 & 12 & 8 & 4 & 6 & 0 \\
			9 & 16 & 6 & 3 & 9 & 0.11 \\
			10 & 16 & 6 & 3 & 10 & 0.20 \\
			11 & 16 & 6 & 3 & 11 & 0.27 \\
		\end{tabular}
	\end{ruledtabular}
	\caption{\label{tab:lat_geom} Lattice geometries used for the finite-size scaling in the bottom panels of (a)~Fig.~\ref{fig:case_study_bosons} and (b)~Fig.~\ref{fig:case_study_fermions}.}
\end{table}

\section{Flux density scaling for Laughlin states on the quartic plane}
\label{sec:add_res}


In this section, we present a total of $1694$ numerical exact diagonalization computations, to complement the results in Sec.~\ref{sec:results}. In particular, we directly follow on from Figs.~\ref{fig:bosons} and~\ref{fig:fermions} and investigate the stability of Laughlin states on the quartic plane as we go deeper into the continuum limit $n_\phi\to 0$.

\begin{figure*}
	\includegraphics[width=\linewidth]{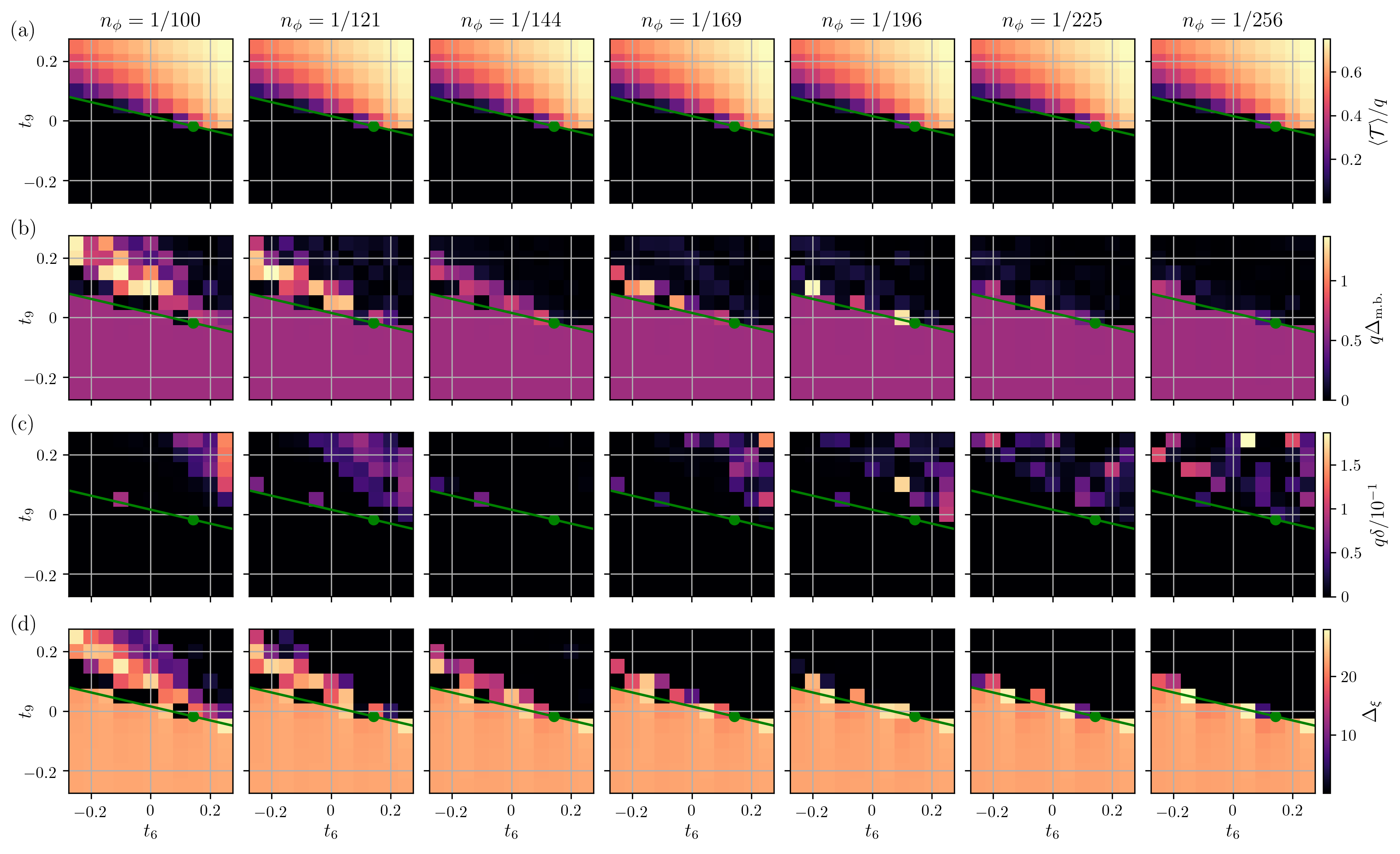}
	\caption{\label{fig:bosons_ext} \textbf{Additional results for the stability of the bosonic Laughlin state on the quartic plane.} (a)~BZ-averaged TISM $\braket{\mathcal{T}}$, scaled by the MUC area $q$, plotted in the quartic plane $1+4t_3+9t_6+16 t_9=0$, for $n_\phi=1/100, 1/121, 1/144, 1/169, 1/196, 1/225, 1/256$. The hexic line $t_9=(1-15t_6)/64$ and octic point $t_9=-1/56$ are overlaid in green. (b)~Many-body gap $\Delta_\text{m.b.}$, scaled by the MUC area $q$, with parameters corresponding to (a). The results are shown for the 8-particle bosonic Laughlin state stabilized by the contact interaction $V_{ij}=\delta_{ij}$. (c)~Quasidegeneracy spread $\delta$, scaled by the MUC area $q$, corresponding to (b). (d)~Principal entanglement gap $\Delta_\xi$, corresponding to (b,c). These results follow directly from Fig.~\ref{fig:bosons}.} 
\end{figure*}

In Fig.~\ref{fig:bosons_ext}, we show a direct continuation of Fig.~\ref{fig:bosons} for the stability of the bosonic Laughlin state at smaller values of $n_\phi$. As in the main text, we consider square configurations with MUCs of dimension $l_x\times l_y = m\times m$ and system dimensions $L_x \times L_y = 4 \times 4$. In the figure, we see that the anomalous regions of large many-body and entanglement gaps above the hexic line gradually diminish as $n_\phi\to 0$, which shows that this is a transient phenomenon. At the smallest flux density of $n_\phi=1/256$, the anomalous regions have almost completely disappeared and the geometric stability hypothesis is recovered, with clear signatures of FCIs observed only below the hexic line, where the TISM is smallest.

\begin{figure*}
	\includegraphics[width=\linewidth]{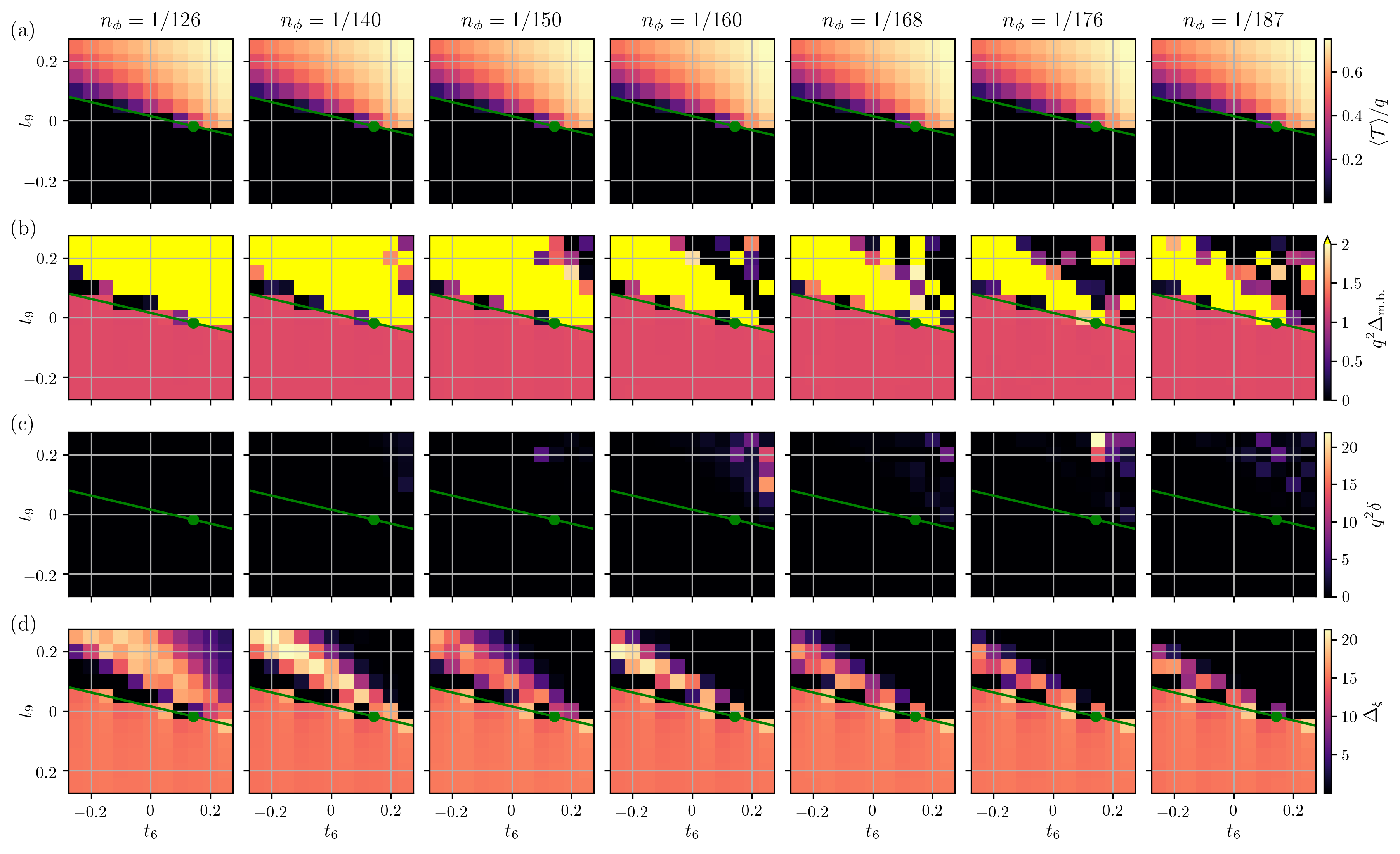}
	\caption{\label{fig:fermions_ext} \textbf{Additional results for the stability of the fermionic Laughlin state on the quartic plane.} (a)~BZ-averaged TISM $\braket{\mathcal{T}}$, scaled by the MUC area $q$, plotted in the quartic plane $1+4t_3+9t_6+16 t_9=0$, for $n_\phi=1/126, 1/140, 1/150, 1/160, 1/168, 1/176, 1/187$. We select flux densities that admit almost square configurations ($\epsilon\leq 7\%$) and are approximately evenly spaced with respect to $q$. The hexic line $t_9=(1-15t_6)/64$ and octic point $t_9=-1/56$ are overlaid in green. (b)~Many-body gap $\Delta_\text{m.b.}$, scaled by the MUC area $q$, with parameters corresponding to (a). The results are shown for the 8-particle fermionic Laughlin state stabilized by the nearest-neighbor interaction $V_{ij}=\delta_{\braket{ij}}$. (c)~Quasidegeneracy spread $\delta$, scaled by the MUC area $q$, corresponding to (b). (d)~Principal entanglement gap $\Delta_\xi$, corresponding to (b,c). These results follow directly from Fig.~\ref{fig:fermions}.} 
\end{figure*}

Similarly, in Fig.~\ref{fig:fermions_ext}, we show a direct continuation of Fig.~\ref{fig:fermions} for the stability of the fermionic Laughlin state at smaller values of $n_\phi$. In this case, we are not able to restrict ourselves to square configurations, due to the increasingly challenging convergence of the Lanczos algorithm. Instead, we consider approximately square configurations with $\epsilon\leq 7\%$, using the algorithm described in Appendix~\ref{sec:num_method}. As for the bosonic Laughlin state in Fig.~\ref{fig:bosons_ext}, we see that the anomalous regions diminish as $n_\phi\to 0$. Similarly, the anomalous regions for the many-body and entanglement gaps are correlated and recede from the top-right corner, where the TISM is largest.

From performing the computations, we find that the anomalous regions are not only fluctuating, but also highly sensitive to the precision of the Lanczos algorithm. Coupled with the fact that these regions are not robust in the thermodynamic or continuum limits, we attribute this to a numerical instability and not a physical phase transition. Since the ratio of the linear system size $\sqrt{N_x N_y} \sim \sqrt{q}$ to magnetic length $\ell \sim (2\pi n_\phi)^{-1/2}\sim\sqrt{q}$ is constant, this is not a length scale phenomenon. Instead, the most likely cause of the transient numerical instability is the relative dominance of the fourth-order momentum term in the Hamiltonian. This highlights the additional care needed to recover the geometric stability hypothesis for Chern bands with a non-Landau level continuum limit.

\bibliography{sfci}

\end{document}